# Physical formula enhanced multi-task learning for pharmacokinetics prediction

Ruifeng Li,[1,2] Dongzhan Zhou,[2] Ancheng Shen,[3] Ao Zhang,[3] Mao Su,[2] Mingqian Li,[4] Hongyang Chen,[4] Gang Chen,[1] Yin Zhang,[*,1] Shufei Zhang,[*,2] Yuqiang Li,[*,2] Wanli Ouyang[2]

[1] College of Computer Science and Technology, Zhejiang University, Hangzhou, Zhejiang, 310027, China.

[2] Shanghai Artificial Intelligence Laboratory, Shanghai, 200232, China.

[3] Shanghai Frontiers Science Center for Drug Target Identification and Delivery, Shanghai Jiao Tong University, Shanghai, 200240, China.

[4] Zhejiang Lab, Hangzhou, Zhejiang, 311100, China.

**Artificial intelligence (AI) technology has demonstrated remarkable potential in drug discovery, where pharmacokinetics plays a crucial role in determining the dosage, safety, and efficacy of new drugs. A major challenge for AI-driven drug discovery (AIDD) is the scarcity of high-quality data, which often requires extensive wet-lab work. A typical example of this is pharmacokinetic experiments. In this work, we develop a physical formula enhanced multi-task learning (PEMAL) method that predicts four key parameters of pharmacokinetics simultaneously. By incorporating physical formulas into the multi-task framework, PEMAL facilitates effective knowledge sharing and target alignment among the pharmacokinetic parameters, thereby enhancing the accuracy of prediction. Our experiments reveal that PEMAL significantly lowers the data demand, compared to typical Graph Neural Networks. Moreover, we demonstrate that PEMAL enhances the robustness to noise, an advantage that conventional Neural Networks do not possess. Another advantage of PEMAL is its high flexibility, which can be potentially applied to other multi-task machine learning scenarios. Overall, our work illustrates the benefits and potential of using PEMAL in AIDD and other scenarios with data scarcity and noise.**

## Introduction

Artificial intelligence-driven drug discovery (AIDD) has demonstrated immense potential in accelerating the drug development process, reducing costs, and enhancing success rates[1-4]. The success of AI models lies in access to a substantial amount of high-quality labeled data. However, the high costs and long periods associated with wet-lab experiments pose significant challenges, leading to data scarcity in AIDD[5,6], as Fig. 1a shows. Furthermore, data noise is an inherent and prevalent issue in wet-lab experiments, with biochemical studies often being especially vulnerable to such inaccuracies and variations[7].

Pharmacokinetics, which delves into the temporal processes and patterns of a drug's absorption, distribution, metabolism, and excretion, is an indispensable component, offering vital guidance to ensure the safety, efficacy, and rational use of medications in the drug discovery process[8-10]. Machine learning[11] has been applied to predict pharmacokinetic parameters[12]. Traditional machine learning approaches often rely on manually designed molecular descriptors[13] or molecular fingerprints[14] and utilize algorithms like Random Forest[15,16], Gaussian Process[17], and Extreme Gradient Boosting[18] for modeling. These methods have low expressiveness and struggle to capture the complex and nonlinear features of molecules that are essential for determining their pharmacokinetic properties. Deep learning[19], on the other hand, can automatically extract key features from the original molecular structures, resulting in semantically richer and more relevant molecular representations[20]. This enables a more effective structure-activity relationship[21] for pharmacokinetic prediction[22,23].

Key parameters of pharmacokinetics include the area under the plasma concentration-time curve (AUC), clearance rate (CL), volume of distribution at steady state (Vdss), and half-life ($T_{1/2}$). Although pharmacokinetic prediction is typically a multi-task problem, existing work often builds separate prediction models for each task. These single-task approaches often struggle to address the data scarcity and noise issues inherent in pharmacokinetics. We propose a multi-task[24] scheme that considers these interrelated parameters simultaneously, enabling the model to capture the underlying mechanism of pharmacokinetics more effectively. Moreover, integrating prior knowledge into neural networks is an effective way to improve the model's accuracy and efficiency, as shown in related work[25-27]. Since the four parameters of pharmacokinetics satisfy two physical formulas, i.e., AUC × CL = $K_1$ and CL × $T_{1/2}$ = Vdss × $K_2$, we can use these formulas to design a multi-task neural network that can predict the four pharmacokinetic parameters. The introduction of physical formulas can serve as explicit constraints among multiple tasks, which facilitates knowledge transfer and target alignment among these tasks. The design leads to a reduced uncertainty of single-task prediction and enhances the accuracy of model predictions. Additionally, the inter-task constraints diminish the model's sensitivity to data noise within a single task.

In this work, we introduce a physical formula enhanced multi-task learning (PEMAL) framework for predicting four pharmacokinetic parameters (Fig. 1b). As pre-training has demonstrated its effectiveness across various tasks[28,29], we adopt a two-stage pre-training strategy within our PEMAL framework. Initially, the model is trained on ample raw molecular structures in a self-supervised manner[30] to learn general molecular representations. Subsequently, we utilize a labeled yet noisy dataset for CL, Vdss, and $T_{1/2}$ to equip our model with task-specific knowledge. It is worth noting that different from other parameters, AUC is heavily influenced by route of administration and dosage, leading to much less data. Therefore, we do not directly pre-train on AUC but rather focus on leveraging the broader insights gained from the other parameters under the multi-task scheme. After pre-training, we integrate the sub-modules into the multi-task system with the guidance of physical formulas. The guidance of physical formulas enables these tasks to align and share knowledge, thus achieving more efficient data utilization and reducing uncertainty. It also strengthens the model's ability to generalize and maintain robustness in various scenarios.

Despite limited data, our model demonstrates remarkable accuracy and generalizability. Trained on only 170 publicly available data points across four pharmacokinetic tasks, PEMAL surpasses both traditional machine learning methods and single-task deep learning methods. Further experimental results indicate that PEMAL exhibits resistance to data noise and maintains high predictive accuracy even with very sparse data. This success highlights the effectiveness of incorporating physical constraints into neural networks, which improves the model's generalizability and robustness, and has special value in scenarios with limited data and high noise. Moreover, two-stage pre-training, using unlabeled structural data and labeled but noisy data, improves the performance of our physically constrained framework. We believe our work has a significant impact on tackling the challenge of scarce and noisy labels, pushing forward the boundaries of multi-task learning, and informing the design of neural networks with a foundation in physical principles.

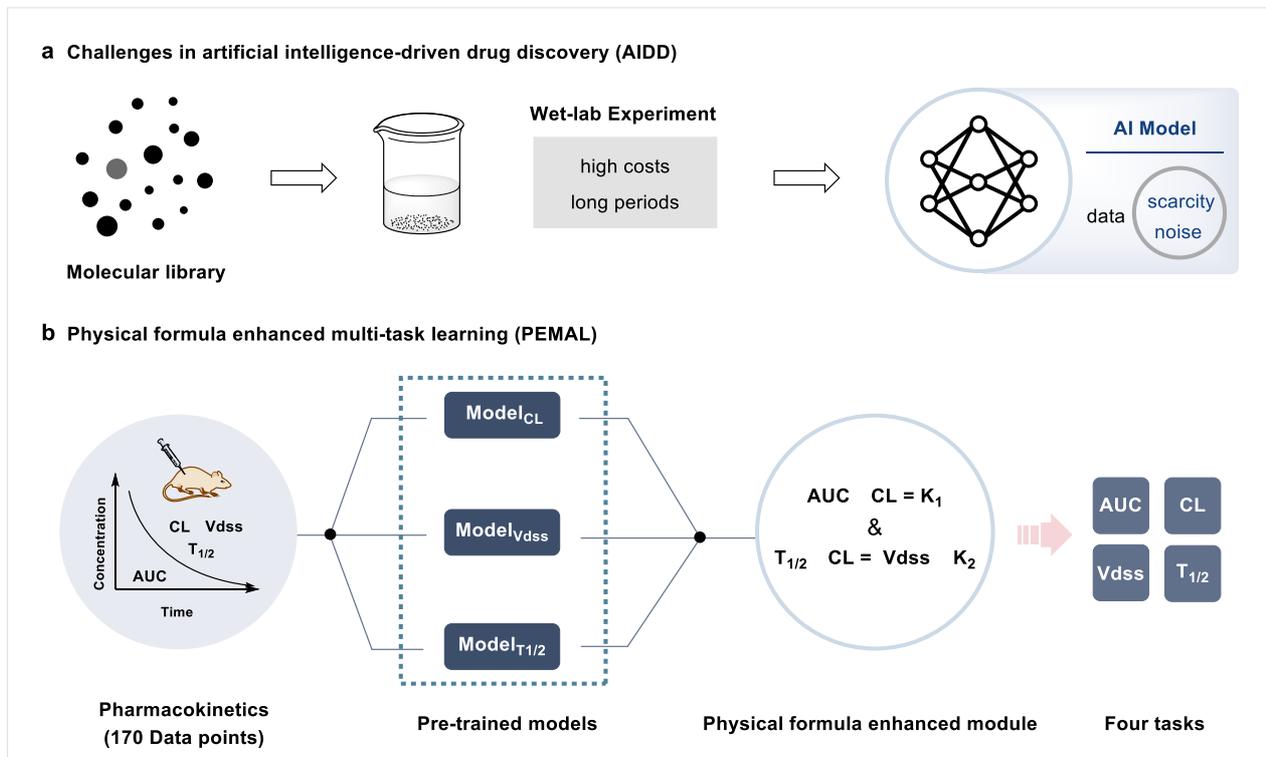

**Fig. 1 Approach overview. a**, Challenges in artificial intelligence-driven drug discovery. **b**, Physical formula enhanced multi-task learning for pharmacokinetics prediction: by incorporating physical formula constraints into the neural network, we enhance the knowledge transfer and target alignment across different tasks.

## Results

### Overview of PEMAL

In this paper, we introduce PEMAL, a multi-task learning framework enhanced by physical formulas for predicting pharmacokinetics. PEMAL comprises three stages: stage I and stage II for pre-training and stage III for multi-task learning with physical formulas. In the pre-training stage, we first utilize self-supervised learning on vast unlabeled molecular data to derive general representations. Next, we perform the second stage of pre-training with labeled but noisy data on the pharmacokinetic tasks (CL, Vdss, and $T_{1/2}$) to inject the task-specific knowledge. In the multi-task learning stage, we integrate physical formulas into the training process to provide explicit constraints and facilitate knowledge sharing.

### Pre-training

**Stage I: Pre-training on unlabeled molecular structures.** By leveraging the ample unlabeled data, self-supervised learning can obtain general representations, benefiting various downstream tasks[30-32]. Currently, there exist two types of paradigms, i.e., contrastive[33,34] and generation-based approaches[35,36]. Existing generation-based methods only consider the information at the atom level but overlook the motif level, which plays an important role in determining molecular properties. Therefore, we devise a dual-level reconstruction task that simultaneously captures atom and motif messages, enabling a more comprehensive understanding of the molecular properties and benefiting the subsequent pharmacokinetic tasks. Additionally, we utilize contrastive learning on top of the dual-level features to further enhance their interactions. The mixture of these two pre-training strategies offers more diverse and robust molecular representations.

As shown in Supplementary Fig. S1, two parallel branches, namely the atom branch and motif branch, are adopted to extract the atom and motif information of the input molecule, respectively. In the atom branch, we first randomly mask several atoms according to a given ratio and then feed the masked molecule into the graph encoder to obtain the atom representations. Before entering the decoder, we re-mask the representations and perform reconstruction based on these corrupted features, which forces the model to learn the intrinsic patterns and improves its generalization ability. The atom branch is restrained by the Mean Absolute Error (MAE) loss between the original molecular graph and the reconstructed graph. In the motif branch, we feed the molecular structure into the graph encoder and derive the motif representations by combining the atom features within each motif. Similar to the atom branch, we perform mask operations on the motif features and require the decoder to reconstruct the representations. The supervision signal comes from the difference between the reconstructed motif representations and the original ones. To facilitate the interactions between the motif level and atom level, we employ a cross-level contrastive learning mechanism, which aims to minimize the distance between representations from the motif and atom branches within the same molecule, while simultaneously maximizing the distance between features of different molecules[29]. Since these two levels of features focus on different aspects, i.e., one on the basic molecular composition and the other on the functional groups of the molecule, contrastive learning contributes to producing more reliable feature representations and offers a more comprehensive understanding.

**Stage II: Pre-training on labeled but noisy pharmacokinetic data.** AUC is the area under the plasma concentration-time curve, which is the core indicator to evaluate the drug absorption degree, reflecting the drug exposure characteristics in the body. However, AUC strongly depends on the dosing method and dosage, resulting in scarce data for specific conditions. CL, Vdss, and $T_{1/2}$ are the drug molecular features defined by AUC, which are less sensitive to the variations of wet-lab experiments, such as dosing method and dosage, which provide a relatively large amount of data across different conditions. To acquire task-specific knowledge for pharmacokinetic parameters, considering the data scarcity of AUC, we further refine the raw structure pre-trained model on tasks of CL, Vdss, and $T_{1/2}$, respectively, whose data are labeled but noisy due to being collected from rats regardless of dosing method and dosage. We expect that the knowledge of AUC can be obtained from the other three parameters by the next stage of multi-task learning constrained by physical formulas. Therefore, we only perform pre-training on CL, Vdss, and $T_{1/2}$ tasks in this stage, where data is more abundant.

As shown in Supplementary Fig. S2, we first load the graph encoder trained in stage I and then attach a fully connected layer for each task to predict the corresponding results. Finally, we compute the MAE loss between the logarithmically transformed predicted values and observed values to supervise the model for each task.

**Physical formula enhanced multi-task learning (Stage III)**

Multi-task learning is an effective strategy to address the issues of data scarcity, as it allows different tasks to share information and complement each other[37,38]. However, existing methods rely on intrinsic information sharing between networks, such as parameter sharing, feature sharing and instance sharing[24]. We argue that these methods are implicit and insufficient to transfer knowledge among tasks that have extrinsic relationships. Therefore, we propose to use physical formulas as extrinsic constraints to connect multiple task modules, which may be a more effective multi-task learning framework for pharmacokinetic prediction. By integrating physical formulas into our training process, we ensure that the knowledge transferred across tasks is not only relevant but also adheres to the physical laws. Moreover, the integration of physical formulas enables the model to resist noise in the data[25]. In this way, we offer a more grounded and principled framework for learning, leading to improved accuracy and robustness in pharmacokinetic prediction.

As shown in Supplementary Fig. S3, we use three parallel branches to predict CL, Vdss, and $T_{1/2}$, respectively, whose weights are inherited from the previous pre-training stages. After getting the CL, Vdss, and $T_{1/2}$, we can infer AUC according to the following equation:

$$AUC \times CL = K_1, \qquad (1)$$

where $K_1$ is a constant. We further require CL, Vdss, and $T_{1/2}$ adhere to the following rule:

$$CL \times T_{1/2} = Vdss \times K_2, \qquad (2)$$

where $K_2$ is also a constant. The supervision signals of the framework originate from two sources: the MAE loss associated with the four parameters (denoted as $L_{AUC}$, $L_{CL}$, $L_{Vdss}$, and $L_{T1/2}$, respectively), and the physical formula constraints in Eq. 2.

**PEMAL boosts the performance of pharmacokinetic prediction**

Pharmacokinetics plays an indispensable role in drug discovery for its crucial involvement in ensuring the efficacy and safety of drugs. Its key parameters include AUC (Area Under the Curve), CL (Clearance), Vdss (Volume of Distribution at Steady State), and $T_{1/2}$ (Half-Life). We have collected and cleaned the dataset named PK-Mol from ChEMBL[39] under the condition of intravenous injection and dosage of 1 mg/kg in rats, which simultaneously includes the four parameters and can be used to systemically evaluate the model performance. We use the MAE between the predicted and ground-truth values as the evaluation metric and all values are logarithmically transformed before calculation.

We evaluate five models on the PK-Mol dataset: (1) Random Forest (RF), (2) Gaussian Process (GP) with handcrafted features, (3) Multilayer Perception (MLP) with handcrafted features, (4) Graph Isomorphism Network (GIN)[40], and (5) PEMAL (ours). The handcrafted features adopted in RF, MLP, and GP are molecular fingerprints MACCS[41] and ECFPs[14].

Table 1 Test performance of different models on PK-Mol dataset[a].

| Method | AUC ↓ | CL ↓ | Vdss ↓ | $T_{1/2}$ ↓ |
|---|---|---|---|---|
| Gaussian Process | 0.913 ± 0.026 | 0.950 ± 0.035 | 0.788 ± 0.014 | 1.431 ± 0.025 |
| Random Forest | 0.984 ± 0.043 | 0.964 ± 0.052 | 0.747 ± 0.011 | 0.864 ± 0.008 |
| MLP | 0.780 ± 0.032 | 0.757 ± 0.042 | 0.768 ± 0.006 | 0.798 ± 0.024 |
| GIN (baseline) | 0.807 ± 0.024 | 0.851 ± 0.030 | 0.753 ± 0.018 | 0.726 ± 0.026 |
| PEMAL | 0.684 ± 0.017 | 0.684 ± 0.016 | 0.664 ± 0.018 | 0.688 ± 0.021 |

a The evaluation metric is MAE, which is calculated based on the logarithmic transformation of the values. The first two models employ machine learning techniques and the others utilize deep learning strategies. We also show the mean discrepancies in MAE over five independent experiments.

As shown in Table 1, PEMAL achieves state-of-the-art (SOTA) performance across all four pharmacokinetic tasks on the test set. Specifically, compared with the most competitive baseline GIN, PEMAL demonstrates MAE reduction of 0.123 (15.2% ↓) for the AUC task, 0.167 (19.6% ↓) for the CL task, 0.089 (11.8% ↓) for the Vdss task, and 0.038 (5.2% ↓) for the $T_{1/2}$ task. The improvements demonstrate the effectiveness of the multi-task scheme and the integration of physical formula constraints.

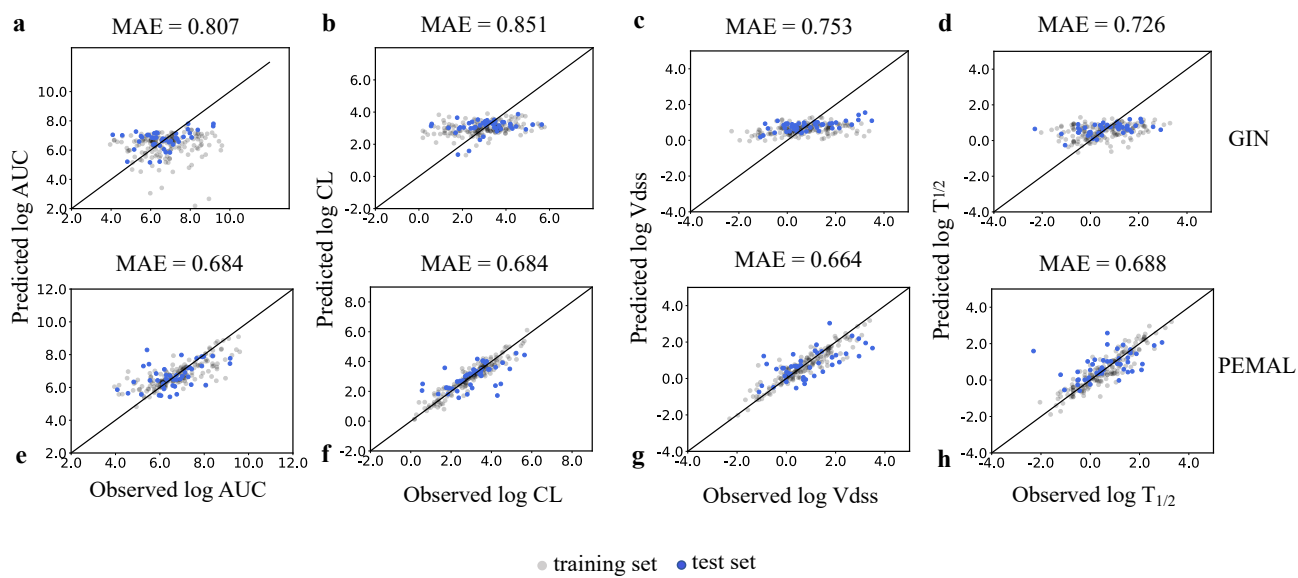

**Fig. 2 Visualization of PEMAL and GIN for pharmacokinetic prediction. a-d** illustrate the predicted and observed values of GIN for AUC, CL, Vdss and $T_{1/2}$, respectively. **e-h** display the predicted and observed values of PEMAL for AUC, CL, Vdss and $T_{1/2}$, respectively. The data from the training set are represented by gray points, while the test set are denoted by blue points. The model performance is better when the points are closer to the diagonal line in the plots.

Fig. 2 provides a detailed visual comparison of model predictions against the experiment-observed value for the pharmacokinetic tasks. Panels a-d show the performance of the GIN (baseline) model, while panels e-h highlight the results of our PEMAL method. These visualizations demonstrate that our PEMAL model achieves a closer alignment between the predicted and actual labels across all four tasks, due to the pretraining and incorporation of physical formula constraints.

**Table 2 Results of the ablation study[a].**

| Index | Stage I | Stage II | Stage III | AUC ↓ | CL ↓ | Vdss ↓ | $T_{1/2}$ ↓ |
|---|---|---|---|---|---|---|---|
| 1[b] | ✗ | ✗ | ✗ | 0.807 | 0.851 | 0.753 | 0.726 |
| 2 | ✓ | ✗ | ✗ | 0.802 | 0.788 | 0.742 | 0721 |
| 3 | ✗ | ✓ | ✗ | - | 0.721 | 0.738 | 0.727 |
| 4 | ✗ | ✗ | ✓ | 0.765 | 0.767 | 0.732 | 0.727 |
| 5 | ✓ | ✓ | ✗ | - | 0.725 | 0.700 | 0.704 |
| 6 | ✓ | ✗ | ✓ | 0.753 | 0.742 | 0.712 | 0.703 |
| 7 | ✗ | ✓ | ✓ | 0.729 | 0.729 | 0.678 | 0.701 |
| 8 | ✓ | ✓ | ✓ | 0.684 | 0.684 | 0.664 | 0.688 |

a MAEs for pharmacokinetics parameters of different stage combinations are reported. Stage I: Pre-training on unlabeled molecular structures. Stage II: Pre-training on labeled but noisy pharmacokinetic data. Stage III: Physical formula enhanced multi-task learning. The results of stage II (row 3) and the stage I+II (row 5) on AUC task are not provided, because the AUC dataset for the stage II is missing. b GIN (baseline).

We conduct extensive ablation experiments to verify the effectiveness of the three stages, and summarize the results in Table 2. The first row presents the results from the GIN model, while row 2 to row 4 depict the outcomes from the individual stages. A comparison between the single-stage results and GIN indicates that each of the three stages enhances PEMAL's performance. Moreover, by comparing row 1 and row 4, we can observe that the reduction in MAE for AUC is the most significant, demonstrating that physical equations can facilitate knowledge transfer and integration from the other three parameters to improve AUC predictions. By inspecting rows 5 to 7, we see that each combination of two stages can further boost performance compared with individual stages. The comparison of results from the combination of stage II+III (row 7) with those from all three stages (row 8) implies that a richer integration of structural information can substantially improve the model's capacity for generalization. The comparison between stage I+III (row 6) and the full stage (row 8) reveals a positive correlation between the noisy pharmacokinetic pre-training and downstream task, that is, the model would perform better if the downstream task aligns with the pre-training task. Additionally, the progression from stages I+II (row 5) to the full stage illustrates the pivotal role of physical formula constraints, which enables a more effective knowledge sharing.

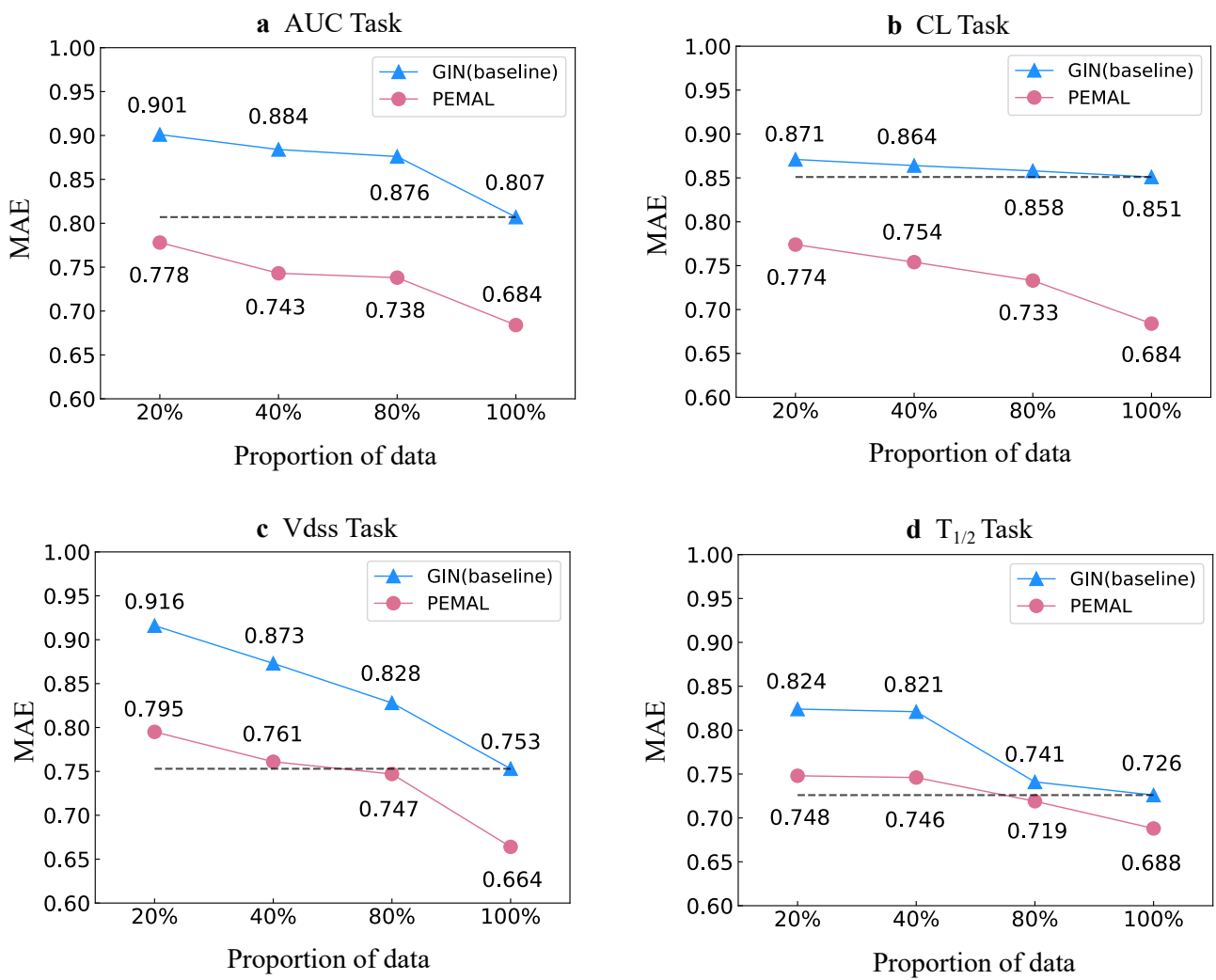

**Fig. 3 Performance across different data volume**. We compare the performance of GIN and PEMAL on **a.** AUC task, **b.** CL task, **c.** Vdss task and **d.** $T_{1/2}$ task with different sizes. The dash line shows the performance of GIN trained with the full dataset.

As data scarcity is a prevalent challenge in AIDD, we explore how varying amounts of data can affect the performance of models. We vary the size of training data through stratified sampling in the PK-Mol dataset and record the MAE for GIN and PEMAL across four pharmacokinetic tasks, as shown in Fig. 3. The results indicate that PEMAL consistently outperforms GIN across all training sizes. Moreover, the advantage of PEMAL over GIN does not decrease with the increase of data volume. For example, in the AUC task (Fig. 3a) with 20 % of the PK-Mol data, PEMAL marks a 13.7% reduction of MAE compared to GIN, from 0.901 to 0.778, and the MAE reduction rate rather increases to 15.2% as the proportion of PK-Mol data increases to 100%. Notably, PEMAL achieves better performance on the AUC (Fig. 3a) and CL (Fig.3b) tasks using only 20% of the training data compared to GIN even using the full dataset. PEMAL consistently demonstrates superior performance to GIN over the other two tasks, even when trained with less data. This impressive results highlight the potential of the multi-task scheme and the incorporation of physical formulations, demonstrating PEMAL's robustness to smaller datasets. We argue that such a capability is especially valuable in pharmacokinetics, where obtaining large amounts of labeled data can be challenging. The success of PEMAL suggests a promising direction for future research in data-efficient machine learning for drug discovery.

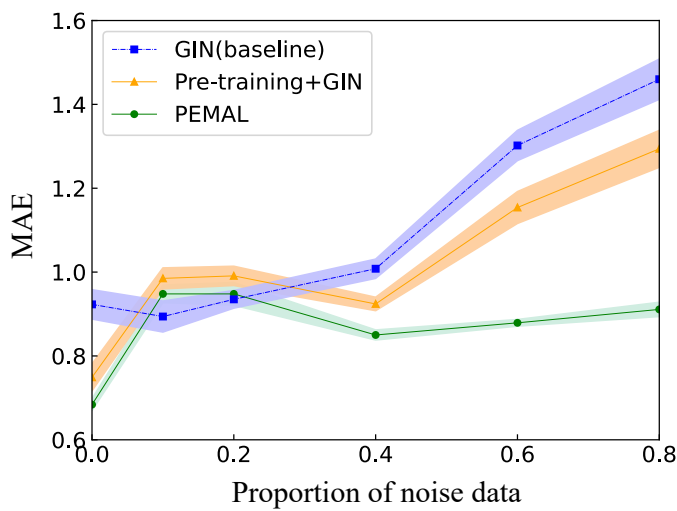

**Fig. 4 Robustness analysis.** In order to evaluate the noise robustness of three methods: GIN, GIN with pre-training (Stage I), and the proposed PEMAL, the proportion of noisy labels in the training set is increased gradually, and the corresponding MAEs' variations are measured.

Most data originate from wet-lab experiments, which poses the data noise as another challenge for AI model training. To assess the robustness of models against this issue, we incrementally introduce noise to the AUC data and monitor its impact on the MAEs of different models, as shown in Fig. 4. The baseline model first encounters an unusual MAE drop with a slight increase in noise (10%), which could be due to regularization and alleviation of over-fitting. But further increase in noise results in a consistent rise in the MAE, revealing its vulnerability. We also observe that pre-training (Stage I) can somewhat alleviate the data scarcity problem but cannot effectively handle the noise issue, as indicated by the drastic increase in MAE when the noise level exceeds 40%. In contrast, PEMAL demonstrates an impressive ability to withstand noise, a resilience attributed to the incorporation of physical formula constraints. This robustness is crucial in the context of pharmacokinetics, indicating PEMAL's advanced capability to deliver favorable pharmacokinetic insights in real-world, noise-affected scenarios.

**Discussion**

The evaluation of pharmacokinetics is crucial for the development of new drugs, but the prediction of pharmacokinetic parameters faces data scarcity and noise challenges, which are common in the whole AIDD field. In this paper, we present PEMAL, a physical formula enhanced multi-task learning (PEMAL) method that predicts four key pharmacokinetic parameters: AUC, CL, Vdss and $T_{1/2}$. Unlike current multi-task learning methods that use internal information sharing, PEMAL uses the physical formulas of the four pharmacokinetic parameters as an external constraint to achieve knowledge sharing and target alignment among multiple tasks, which can reduce the model's data requirement and resist data noise.

Our results show that PEMAL outperforms traditional machine learning methods, such as Gaussian Process (GP) and Random Forest (RF), and deep learning methods, such as Multilayer Perception (MLP) and Graph

Isomorphism Network (GIN), on all four pharmacokinetic tasks. We also show that the two-stage pre-training and the incorporation of physical formulas are crucial components of PEMAL, as they improve the transfer and enrichment of knowledge among the pharmacokinetic parameters, resulting in higher prediction accuracy for all tasks. Furthermore, we conduct data volume analysis to show that PEMAL significantly reduces the data requirement compared with GIN. Additionally, robustness experiments show the strong ability of the physical formula-based network to resist data noise that cannot be solved by pre-training methods.

It is worth noting that although we also use the pre-training method in PEMAL, our physical formula framework does not depend on a specific pre-training method and a specific backbone for a single task, which indicates that PEMAL has a wide range of transferability and has the potential to be applied in other multi-task scenarios with physical formula constraints.

In conclusion, we propose PEMAL, a novel approach that leverages physical formulas to design neural networks for multiple tasks. By imposing explicit constraints, PEMAL enables knowledge sharing and target alignment among different tasks, leading to superior performance, robustness to data noise, and broad transferability. These advantages make PEMAL a promising method for AIDD and other scenarios that face challenges of data scarcity and noise. We hope that our work can stimulate more research on multi-task learning and physical knowledge integration for AIDD and beyond.

## Methods

In this section, we introduce the methodology employed in our work. We begin by showing the detailed implementation of our PEMAL framework, which includes two-stage pre-training and multi-task learning enhanced with the physical formulas. Following that, we describe the dataset, preprocessing steps, and other training configurations.

## PEMAL network

**Stage I: Pre-training on unlabeled molecular structures.** As shown in Supplementary Fig. S1, we perform the self-supervised pre-training on both the atom and motif levels, where the supervision signal comes from a mixture of the generation-based loss and the contrastive loss. The input can be denoted as $\mathcal{G} = (\mathcal{V}, \mathcal{U}, A_a, A_m, X)$, where the subscripts $a$ and $m$ refer to the atom level and the motif level, respectively. $\mathcal{V}$ represents the node set, while $\mathcal{U}$ indicates the specific motif associated with each node. $N = |\mathcal{V}|$ is the number of nodes and $X \in \mathbb{R}^{N \times d_{init}}$ stands for the matrix of node features. $A_a \in \{0,1\}^{N \times N}$ represents the adjacency matrix of node. $A_m \in \{0,1\}^{C \times C}$ refers to the adjacency matrix of motif, where $C$ is the number of motifs.

At the atom branch, a part of the atoms is masked before passing through the encoder. These masked atoms are randomly selected based on a uniform distribution, denoted as $\tilde{\mathcal{V}} \in \mathcal{V}$, whose features are replaced by a mask token [MASK]. Thus, the corrupted feature matrix $\tilde{X} \in \mathbb{R}^{N \times d_{\text{init}}}$ can be written as follows:

$$\tilde{x}_i = \begin{cases} x_{[MASK]} & v_i \in \tilde{\mathcal{V}} \\ x_i & v_i \notin \tilde{\mathcal{V}} \end{cases}, \tag{3}$$

where $\tilde{x}_i$ is the $i^{th}$ atom feature in corrupted feature matrix $\tilde{X}$.

Then the original graph $(A, X)$ and corrupted graph $(A, \tilde{X})$ are fed into the encoder $f_E$:

$$H_a = f_E(A_a, \tilde{X}), \quad H_m = f_E(A, X), \tag{4}$$

where $H_a \in \mathbb{R}^{N \times d}$ represents the atom representation matrix and $H_m \in \mathbb{R}^{C \times d}$ is the motif representation matrix. Specifically, the motif representations are obtained by pooling the representations of atoms within the motif.

To encourage the encoder to learn more generalized and unified representations, we perform a masking operation on $H_a$ and $H_m$ to obtain masked atom representations $\tilde{H}_a$ and motif representations $\tilde{H}_m$. The corresponding feature is replaced by mask token [DAMASK] and [DMMASK], respectively, as shown below:

$$\tilde{h}_{a,i} = \begin{cases} h_{a,[DAMASK]} & v_i \in \tilde{\mathcal{V}} \\ h_{a,i} & v_i \notin \tilde{\mathcal{V}} \end{cases}, \quad \tilde{h}_{m,i} = \begin{cases} h_{m,[DMMASK]} & u_i \in \tilde{\mathcal{U}} \\ h_{m,i} & u_i \notin \tilde{\mathcal{U}} \end{cases}, \tag{5}$$

where $h_{a,i}$ represents the $i^{th}$ feature of atom representation matrix $H_a$, $h_{m,j}$ is the $j^{th}$ feature of motif representation matrix $H_m$, $\tilde{h}_{a,i}$ denotes the $i^{th}$ masked atom representation and $\tilde{h}_{m,j}$ is the $j^{th}$ masked motif representation. $\tilde{\mathcal{U}}$ represents the subset of motifs $\mathcal{U}$.

Afterward, the masked atom feature matrix $\tilde{H}_a$ and the masked motif featurematrix $\tilde{H}_m$ pass through graph decoders to reconstruct the atom feature matrix $Z_a$ and motif feature matrix $Z_m$, respectively:

$$Z_a = f_D^a(A_a, \tilde{H}_a), \quad Z_m = f_D^m(A_m, \tilde{H}_m), \tag{6}$$

where $f_D^a$ and $f_D^m$ refer to the decoder for atoms and motifs, respectively.

We adopt the scaled cosine error (SCE) as the reconstruction loss between the ground-truth and masked features, as depicted in Eq. 7 :

$$\mathcal{L}_{a,SCE} = \mathbb{E}_{v_i \in \tilde{\mathcal{V}}}(1 - x_i^\top z_{a,i})^{\gamma_a}, \quad \mathcal{L}_{m,SCE} = \mathbb{E}_{u_j \in \tilde{\mathcal{U}}}(1 - h_{m,j}^\top z_{m,j})^{\gamma_m}, \tag{7}$$

where scaling factors $\gamma_a > 1$ and $\gamma_m > 1$ are the hyperparameters to adjust the weight of each sample based on the reconstruction error. $x_i$ denotes the $i^{th}$ atom ground-truth feature and $h_{m,j}$ is the $j^{th}$ motif ground-truth feature. $z_{a,i}$ represents the $i^{th}$ atom reconstructed feature and $z_{m,j}$ is the $j^{th}$ motif reconstructed feature.

To integrate the structural information at the atom and fragment levels more effectively, we design a contrastive loss to align the cross-level molecular representations. In this way, we obtain more versatile and semantically enriched molecular representations. The atom-level molecular representations $P_a$ and the motif-level representations $P_m$ are acquired through the READOUT operation, that is,

$$P_a = READOUT(H_a), \quad P_m = READOUT(H_m), \tag{8}$$

where we use the sum pooling as the READOUT.

Subsequently, we adopt a contrastive loss[34,42] to align the molecular representations at different levels,

$$\mathcal{L}_{cons} = -\frac{1}{B}\sum_{i=1}^{B} \log \frac{exp(sim(\boldsymbol{P}_{a,i}, \boldsymbol{P}_{m,i})/\tau)}{\sum_{k=1}^{2B} \mathbb{1}\{k \neq i\} exp(sim(\boldsymbol{P}_{a,i}, \boldsymbol{P}_{m,k})/\tau)}, \tag{9}$$

where $B$ is the batch size, sim(·) measures the similarity between different levels of molecular representations and $\tau$ is the temperature parameter.

The overall loss function for the first stage pre-training is shown as follows:

$$\mathcal{L} = \mathcal{L}_{a,SCE} + \alpha \mathcal{L}_{m,SCE} + \beta \mathcal{L}_{cons}, \tag{10}$$

where $\alpha$ and $\beta$ are hyperparameters to adjust the weight of atom reconstruction loss $\mathcal{L}_{a,SCE}$ and motif reconstruction loss $\mathcal{L}_{m,SCE}$, respectively. $\mathcal{L}_{cons}$ is the contrastive loss of atom level and motif level molecular representations.

**Stage II: Pre-training on labeled but noisy pharmacokinetic data.** As shown in Supplementary Fig. S2, in the second stage of pre-training, we use the encoder from the first stage as the backbone and append fully connected layer to predict the pharmacokinetic parameters CL, Vdss, and $T_{1/2}$, thereby acquiring domain knowledge from three datasets. The networks trained in stage II are saved as CL model $f_c$, Vdss model $f_v$, and $T_{1/2}$ model $f_t$, respectively.

**Stage III: Physical formula enhanced multi-task learning.** As shown in Supplementary Fig. S3, predicting pharmacokinetics is naturally multi-objective as its key parameters adhere to the laws of biochemistry. We use physical formulas to describe these relationships and supervise the multi-task learning scheme to facilitate knowledge transfer. The three models that contain task-specific knowledge (from stage II) are integrated into the whole system and generate predictions for CL, Vdss, and $T_{1/2}$, respectively, as shown below:

$$y_{CL} = f_c(\boldsymbol{A}_a, \boldsymbol{X}), \quad y_{Vdss} = f_v(\boldsymbol{A}_a, \boldsymbol{X}), \quad y_{T_{1/2}} = f_t(\boldsymbol{A}_a, \boldsymbol{X}), \tag{11}$$

where $y_i$ ($i \in \{CL, Vdss, T_{1/2}\}$) is the predicted value.

Based on the constraints of the physical formula 1 (Eq. 1), we can calculate the AUC prediction in the following way,

$$y_{AUC} = \boldsymbol{K}_1 / y_{CL} \tag{12}$$

We require the other three parameters to satisfy the constraints of formula 2 (Eq. 2):

$$\mathcal{L}_{formula} = \left| y_{CL} \cdot y_{T_{1/2}} / y_{Vdss} - \boldsymbol{K}_2 \right| \tag{13}$$

**Experimental Setup**

**Pre-training and downstream dataset.** In the raw molecular structure pre-training stage, a substantial unlabeled dataset comprising 904,816 molecules is amassed from the publicly accessible ChEMBL[39] database. Given that ChEMBL primarily catalogs molecules encountered in pharmaceutical research, we assume that the structural distribution of our unlabeled dataset closely reflects the compounds pertinent to pharmacokinetics.

Pre-training on this large scale dataset enables a generalized representation learning of drug molecules, which is essential for robust prediction of pharmacokinetics.

In the noisy pharmacokinetic pre-training stage, the CL, Vdss, and $T_{1/2}$ parameters are less influenced by the experimental conditions, such as the dosing method and dosage, leading to a relatively large dataset. We collect 9,982, 4,771, and 10,126 molecules for CL, Vdss, and $T_{1/2}$, respectively. The animal type is fixed to rats for all the tasks.

The CL and $T_{1/2}$ datasets are proportionally divided into training, validation, and test sets in a 7:1:2 ratio, while the Vdss dataset is divided into an 8:1:1 ratio due to its smaller size.

In the physical formula enhanced multi-task learning stage, we collect the dataset, PK-Mol, which includes labels for all four pharmacokinetic parameters. The data selection follows specific criteria: only instances with an animal model that is a rat, with administration by injection, and with a dosage of 1 mg/kg are included. This intersecting dataset is then used to systemically evaluate the efficacy of models. The data are proportionally divided into training, validation, and test sets in a 7:1:2 ratio.

**Implementation details.** Raw molecular data are stored as Simplified Molecular Input Line Entry System (SMILES) strings[43]. To process the data, we utilize the widely used RDKit library[44] to convert SMILES into two-dimensional molecular graphs. Features of atoms include atom type and chirality, while features of bonds comprise bond type and direction. The BRICS algorithm [45] within the RDKit package is adopted for molecular decomposition.

During the raw molecular structure pre-training stage, the learning rate is set at 0.001, with a batch size of 256. The optimizer is Adam[46] and the weight decay is 0. The similarity loss is a scaled cosine error. GIN[40] serves as the backbone network and the dimension of all embeddings is set to 300. The masking ratios for atoms and fragments are 0.75 and 0.25, respectively, while alpha and beta parameters are both set at 0.5.

In the noisy pharmacokinetic pre-training stage, we train three separate models for the CL, Vdss, and $T_{1/2}$ datasets, respectively. The three models have the same network architecture, which is GIN followed by a fully connected layer. The optimizer used for the three models is Adam, with a batch size of 64. The loss function is MAE loss. The learning rate is set to 0.0001 for CL and $T_{1/2}$, while for Vdss, it is 0.0005.

In the physical formula enhanced multi-task learning stage, the learning rate is set to 0.0001, with a batch size of 32 and a weight decay of 1e-9. We use Adam[46] to optimize the framework. We perform a grid search for hyperparameter selection. The framework is implemented via PyTorch on a single NVIDIA A100 GPU. More details can be seen in Supplementary Material.

**Data availability:** The data used for pre-training and pharmacokinetic tasks are available at https://drive.google.com/drive/folders/1zYvoPbI5pS8S1dPdX5kBbAk6WBHbxmaR?usp=drive_link.

**Code availability:** The source code for reproducing the findings in this paper are available at https://github.com/Lirain21/PEMAL.git.

**Acknowledgments:** This work is partially supported by the National Key R&D Program of China (NO.2022ZD0160101, NO.2022YFB4500300), National Natural Science Foundation of China under Grant 62271452, Zhejiang Provincial Natural Science Foundation of China under Grant No. LZ23F020009, the NSFC project (No. 62072399), and the Fundamental Research Funds for the Central Universities. This work was done during Ruifeng Li's internship at Shanghai Artificial Intelligence Laboratory.

**Author contributions:** Y.L. designed the project and directed the work. R.L. and S.Z. developed the codes and trained the models. Y.L., R.L. D.Z. and S.Z. wrote the manuscript. All authors contributed to shape the research, provided critical feedback, and commented on the paper and its revisions.

**Competing interests:** Authors declare no competing interests.

Material & Correspondence: Dr. Yuqiang Li, liyuqiang@pjlab.org.cn, Dr. Shufei Zhang, zhangshufei@pjlab.org.cn. Prof. Yin Zhang, zhangyin98@zju.edu.cn.

# Supplementary Materials

## PEMAL Details

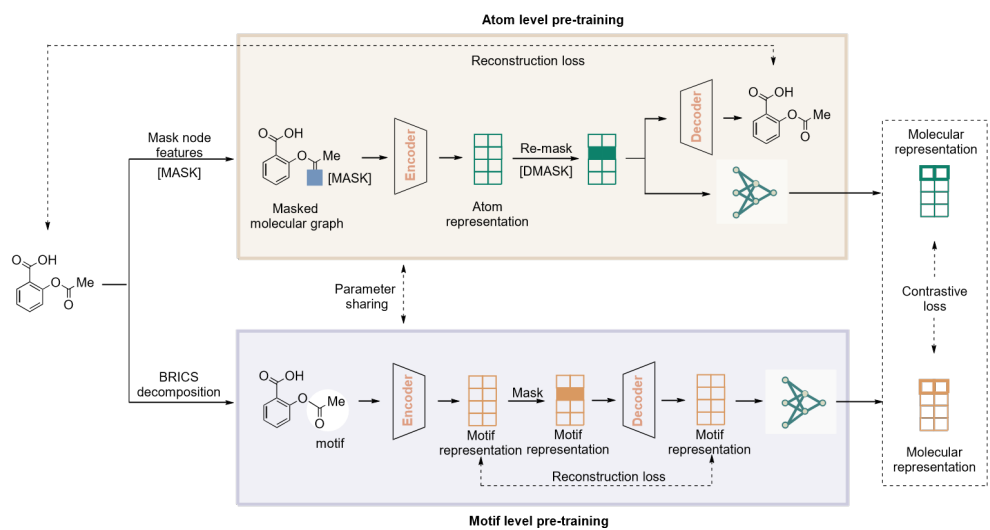

**Fig. S1 Stage I: Pre-training on unlabeled molecular structures.** The structure information on atom-level and motif-level is extracted through dual-level masking scheme. Additionally, we employ contrastive learning on dual-level features to further improve the diversity and robustness of molecular representations.

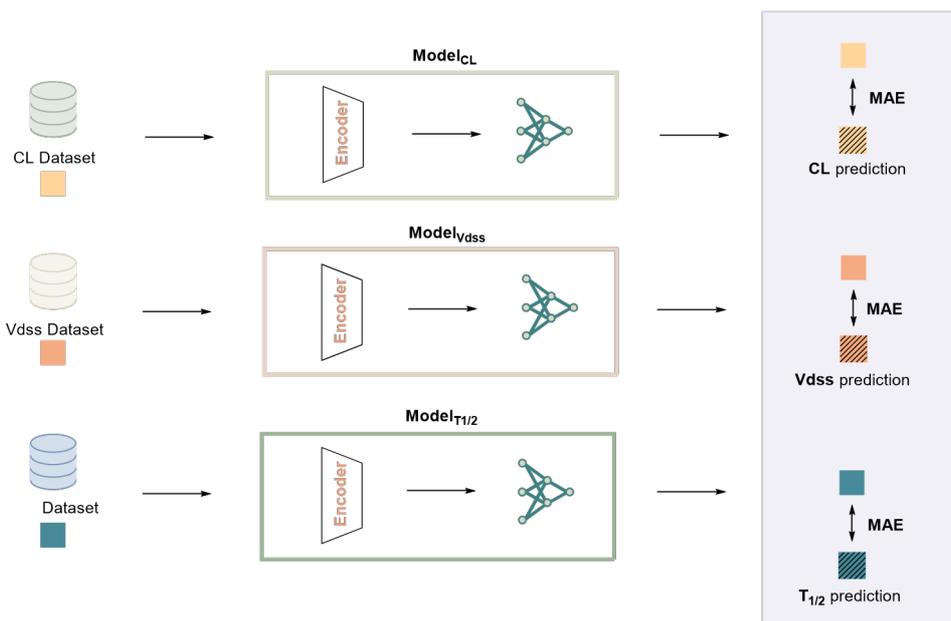

**Fig. S2 Stage II: Pre-training on labeled but noisy pharmacokinetic data.** We perform pre-training on three pharmacokinetic tasks, namely CL, Vdss, and $T_{1/2}$. We use the graph encoder from Stage I as the backbone for the three tasks.

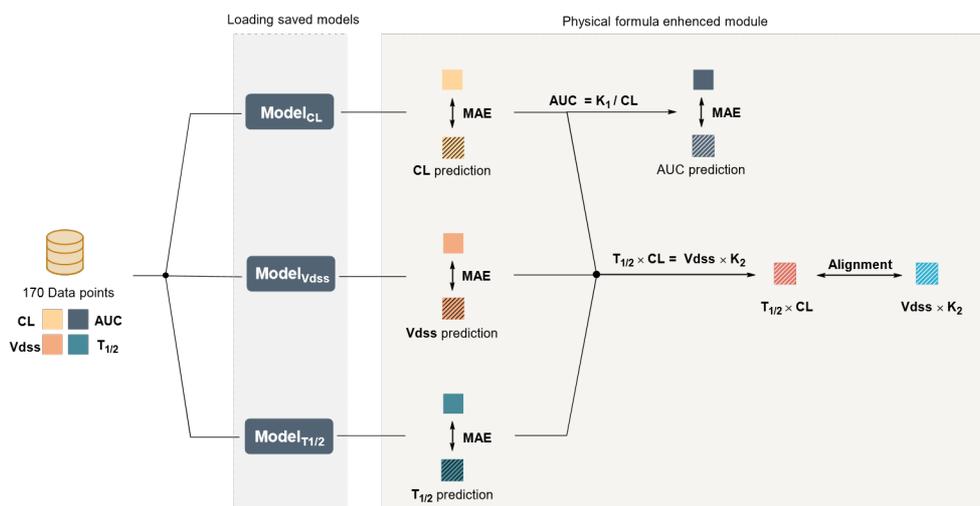

**Fig. S3 Stage III: Physical formula enhanced multi-task learning.** By incorporating physical formula constraints into the neural network, we enhance the knowledge transfer and objective alignment across different tasks.

## Dataset Details

We select data of rats from the ChEMBL database[1] to build our dataset for the pharmacokinetic parameters. In stage III, since the AUC task is particularly susceptible to experimental conditions, we add two additional constraints besides animal type, that is, intravenous injection as the administration method and 1mg/kg as the dosage. Finally, we obtain 215 data points that have labels for all pharmacokinetic parameters.

Data volume and screening criteria for each stage are summarized in Table 1.

**Table 1 Dataset details for each stage[a].**

| Stage | Data volume | Label type |
| --- | --- | --- |
| Stage I | 904,816 | - |
| Stage II | 9,982 | CL |
|  | 4,771 | Vdss |
|  | 10,126 | $T_{1/2}$ |
| Stage III | 215 | AUC, CL, Vdss and $T_{1/2}$ |

a All dataset are collected from ChEMBL database. In the stage I, we filter data based on IC50; in the stage II, we filter data based on rat; and in the stage III, we filter data based on "intravenous injection at a concentration of 1 mg/kg in rats."

## Baseline methods

We compare PEMAL with four other methods: (1) Gaussian Process (GP)[2] (2) Random Forest (RF)[3], (3) Multilayer Perception (MLP), (4) Graph Isomorphism Network (GIN)[4] GP is a commonly applied non-parametric Bayesian model for regression and classification tasks, which is capable of providing uncertainty estimates for predictions. RF is an ensemble learning method for regression by constructing a number of decision trees. We directly invoke the integrated scikit-learn[5] to implement GP and RF. MLP, as one of the simplest deep neural networks, possesses a relatively strong expressive capability. GIN excels in graph representation learning, particularly in tasks such as graph classification and isomorphism testing, where the complex relationships between graph topology and node features can be effectively learned.

**The PEMAL: Details and hyperparameters**

In Table 2, we show the hyperparameters for the pre-training on stage I. The configurations of CL, Vdss, and $T_{1/2}$ for stage II are shown in Table 3, Table 4, and Table 5, respectively. The hyperparameters of stage III are summarized in Table 6.

Table 2 Hyperparameter space considered for the PEMAL with the pre-training on stage I.

| Hyperparameter | Explored values |
| --- | --- |
| Nmber of GNN layers | 5 |
| Feature dimension | 300 |
| Batch size | 64 |
| Learning rate | 0.001 |
| Optimizer | Adam |
| Weight decay | 0 |
| Dropout | 0.0 |
| Layer-normalization | False |
| Reconstruction loss function | Scaled cosine error |
| Constrastive loss function | InfoNCE loss |
| Atom mask ratio | 0.75 |
| Motif mask ratio | 0.25 |
| Mask SCE ratio | 0.5 |
| Contrastive ratio | 0.5 |

**Table 3 Hyperparameter space considered for the CL model with the pre-training on stage II.**

| Hyperparameter | Explored values |
|---|---|
| Nmber of GNN layers | 5 |
| Feature dimension | 300 |
| Batch size | 64 |
| Learning rate | 0.0001 |
| Optimizer | Adam |
| Weight decay | 1e-10 |
| Dropout | 0.0 |
| Layer-normalization | True |

**Table 4 Hyperparameter space considered for the Vdss model with the pre-training on stage II.**

| Hyperparameter | Explored values |
|---|---|
| Nmber of GNN layers | 5 |
| Feature dimension | 300 |
| Batch size | 64 |
| Learning rate | 0.0005 |
| Optimizer | Adam |
| Weight decay | 1e-10 |
| Dropout | 0.0 |
| Layer-normalization | True |

**Table 5 Hyperparameter space considered for the $T_{1/2}$ model with the pre-training on stage II.**

| Hyperparameter | Explored values |
| --- | --- |
| Nmber of GNN layers | 5 |
| Feature dimension | 300 |
| Batch size | 64 |
| Learning rate | 0.0001 |
| Optimizer | Adam |
| Weight decay | 1e-10 |
| Dropout | 0.0 |
| Layer-normalization | True |

**Table 6 Hyperparameter space with the stage III: physical formula enhanced multi-task learning.**

| Hyperparameter | Explored values |
| --- | --- |
| Nmber of GNN layers | 5 |
| Feature dimension | 300 |
| Batch size | 32 |
| Learning rate | 0.0001 |
| Optimizer | Adam |
| Weight decay | 1e-9 |
| Dropout | 0.0 |
| Layer-normalization | True |
| $K_1$ | 16,846 |
| $K_2$ | 17.71 |
| CL learning rate scale | 1.0 |
| Vdss learning rate scale | 1.0 |
| $T_{1/2}$ learning rate scale | 1.0 ~ 2.2 |